%% file: paper_rereviewed.tex
\documentclass[usegraphix,usenatbib]{mn2e}

\include{defn}

\usepackage{amsmath}
\usepackage{amssymb}
\usepackage{times}
\usepackage{graphicx}
\usepackage{subfigure}
\usepackage{epsfig}
\usepackage{txfonts}
\usepackage{epstopdf}
\usepackage{gensymb}

\def\vlos{{v_\mathrm{LOS}}}
\def\intd{{\mathrm{d}}}

\def\acts{{\boldsymbol{J}}}

\def\kms{{\,\mathrm{kms^{-1}}}}
\def\kpc{{\,\mathrm{kpc}}}

\def\rmpi{{\mathrm{\pi}}}
\def\JB{{J_\mathrm{b}}}
\def\ain{{\alpha_\mathrm{in}}}
\def\aout{{\alpha_\mathrm{out}}}
\def\rB{{r_\mathrm{b}}}

\def\a0{{\alpha_0}}
\def\ainf{{\alpha_\infty}}

\begin{document}

\pagenumbering{arabic}

\title[Dynamical Stellar Halo Models]
  {Haloes light and dark: dynamical models of the stellar halo and constraints on the mass of the Galaxy}

\author[Williams \& Evans]
  {A.A. Williams$^1$\thanks{E-mail: aamw3,nwe@ast.cam.ac.uk},
   N.W. Evans$^1$
 \medskip
 \\$^1$Institute of Astronomy, University of Cambridge, Madingley Road,
       Cambridge, CB3 0HA, UK}

\maketitle

\begin{abstract}
We develop a flexible set of action-based distribution functions (DFs)
for stellar haloes. The DFs have five free parameters, controlling the
inner and outer density slope, break radius, flattening and anisotropy
respectively. The DFs generate flattened stellar haloes with a rapidly
varying logarithmic slope in density, as well as a spherically aligned
velocity ellipsoid with a long axis that points towards the Galactic
centre -- all attributes possessed by the stellar halo of the Milky
Way.  We use our action--based distribution function to model the blue
horizontal branch stars extracted from the {\it Sloan Digital Sky
  Survey} as stellar halo tracers in a spherical Galactic potential.
As the selection function is hard to model, we fix the density law
from earlier studies and solve for the anisotropy and gravitational
potential parameters. Our best fit model has a velocity anisotropy
that becomes more radially anisotropic on moving outwards. It changes
from $\beta \approx 0.4$ at Galactocentric radius of 15 kpc to
$\approx 0.7$ at 60 kpc. This is a gentler increase than is typically
found in simulations of stellar haloes built from the mutiple
accretion of smaller systems. We find the potential corresponds to an
almost flat rotation curve with amplitude of $\approx 200$ kms$^{-1}$
at these distances. This implies an enclosed mass of $\approx 4.5
\times 10^{11} M_\odot$ within a spherical shell of radius $50\kpc$.
\end{abstract}


\section{INTRODUCTION}

It has long been known that the distribution functions (DFs) of
collisionless stellar systems depend on the integrals of motion. This
result is often called \citet{Je19} Theorem, though it was known
earlier to Poincar\'e. A number of choices of integrals of motion are
possible. These include the classical integrals such as energy or
angular momentum~\citep[e.g.,][]{Ed15, Hu93, Ev06}, the turning points
or apocentric and pericentric distances of orbits~\citep{Hu92},
or the actions~\citep[e.g.,][]{Bi84,Bi87}.  Although there are
advantages and disadvantages to any of these choices, the actions
remain very appealing because they are adiabatically invariant and
they form a canonical set of coordinates when combined with the
conjugate angles~\citep[see e.g.,][]{Go80}.

Historically, actions have not been much used in stellar dynamics
because of the difficulty in calculating them. For classical problems
like the Keplerian potential or the harmonic oscillator, they can be
readily evaluated by residue calculus~\citep{Bo27}. The isochrone is
the most general spherical potential for which the actions are
available as elementary functions~\citep[e.g.,][]{He59,Ev90}.  For the
separable or St\"ackel potentials, they are available as a quadrature
and provide a beautiful demarcation of phase space in terms of the
orbital types~\citep{deZ85}.  However, the last few years have seen a
major change, in that a number of methods have been proposed to enable
the rapid evaluation of actions in a variety of potentials. This
includes the adiabatic approximation~\citep{Bi10}, local
fitting~\citep{Sa12} and the St\"ackel
fudge~\citep{Bi12a,Sa15a}. Together with the release of software
packages like {\sc galpy}~\citep{Bov15}, this work has transformed the
calculation of actions into a routine matter for most axisymmetric and
triaxial potentials.

The form of the DF in action space is now an interesting problem to
study. The complex nature of galaxies suggests that each component --
bulge, disk, stellar halo and dark halo --- needs its own distribution
function and is described by a characteristic shape in action
space~\citep{Bi87}. The functional form of the DF for thin and thick
stellar disks is suggested by warming up models based on pure circular 
or epicyclic orbits and has been studied extensively in recent years,
motivated by applications to the Milky Way~\citep[see e.g.,][]{Bi10,
  Bo13}. \citet{Bi12b} showed that action-based DFs fit to the
Geneva-Copenhagen Survey of nearby stars~\citep{No04} accurately
predict the kinematics of stars in the deeper Radial velocity
experiment~\citep[RAVE,][]{St06}. Subsequently, \citet{Sa15b}
introduced extended DFs for the Galactic disk by introducing an
analytic dedendence on metallicity. However, hot components such as
bulges, dark matter haloes and stellar haloes have received much less
attention.  \cite{Po15} and \cite{Wi15} considered self--consistent
models with double power-law density profiles. In principle, such DFs
can be used for stellar and dark haloes, as well as elliptical and
dwarf spheroidal galaxies. However, these papers did not provide any
fits to large datasets.  We aim to remedy this deficiency here, by
using such models to represent the Milky Way's stellar halo as a
tracer population in the Galactic potential.

Section 2 introduces our family of DFs which depend on the
actions. These are motivated by the density law of the stellar halo,
which is of a flattened double power-law form to a reasonable
approximation. Section 3 provides a brief description of the data on
blue horizontal branch stars extracted by \citet{Xu11} from the Sloan
Digital Sky Survey. As the selection function of the sample is
unknowable, Section 4 outlines our fitting methodology.  Section 5
presents our results, together with a comparison with other recent
work on the Milky Way stellar halo. We sum up in Section 6, and
suggest further extensions and applications of our models.

\section{The dynamical model}

In this section, we describe the construction of our dynamical
model. First, we summarise the observational constraints already
placed on the properties of the distribution function. We then
demonstrate how these constraints can be used to construct a suitable
model.

\subsection{Observational motivation}
\label{sec:obsmot}

The spatial properties of the stellar halo are reasonably
well-studied. \citet{De11c} found that a sample of $\sim 20000$ Blue Horizontal Branch (BHB)
and Blue Straggler (BS) stars taken from the Sloan Digital Sky Survey (SDSS) is well
modelled by a flattened broken power law
\begin{equation}
\rho(\boldsymbol{r}) \propto \begin{cases}
\left(\dfrac{r_q}{\rB}\right)^{-\ain} & r_q < \rB, \\ \\
\left(\dfrac{r_q}{\rB}\right)^{-\aout} & r_q > \rB, \end{cases}
\end{equation} 
where $r_q^2 = R^2 + (z/q)^2$. Their best-fit parameters are $\rB = 27
\kpc$, $\ain = 2.3$, $\aout = 4.6$ and $q = 0.59$. Other studies have
provided similar, though not identical, results \citep[see
  e.g.,][]{Pil15} and so we can be confident that the stellar-halo
density is well-represented by oblate models with a sharp change in
gradient in the density profile at $\sim 30 \kpc$. This rapidly
changing gradient in the density is equally well described by the
Einasto density law
\begin{equation}
\rho(r_q) \propto \exp\left( -d_n \left[\left(r_q/r_\mathrm{eff}\right)^{1/n} - 1\right]\right),
\end{equation}
where $d_n = 3n - 1/3 + 0.0079/n$.  \citet{De11c} also fit such a
profile in their analysis, with best fit parameters are
$\left(n,q,r_\mathrm{eff}\right) = \left(1.7,0.58,20\kpc\right)$.

The structure of halo stars in velocity space is not as well
constrained, which is primarily explained by the comparative lack of
information: we possess all three spatial coordinates, but usually only one
component of the velocity (the line--of--sight velocity), of each halo
star. Nonetheless, the fact that the sun resides at a distance from
the Galactic Centre still allows us to place constraints on the
velocity ellipsoid of the stellar halo. \citet{Bo10} used a large
sample of SDSS halo stars to investigate the velocity structure within
$r_\mathrm{GC} \sim 10 \kpc$, and found that the velocity ellipsoid is
essentially spatially invariant within their distance limit. Their
inferred ellipsoid aligns with spherical coordinates, with axes given
by $\sigma _r = 141 \kms$, $\sigma _\phi = 85 \kms$ and $\sigma
_\theta = 75 \kms$. The spherical anisotropy parameter, given by
\begin{equation}
\beta(r) = 1 - \dfrac{\sigma _\phi ^2 + \sigma _\theta ^2}{2\sigma _r ^2},
\end{equation}
has a value of $\sim 0.65$. Other studies of halo kinematics find
similar results, with $\beta$ usually between 0.4 and
0.6~\citep[e.g.,][]{Sm09a}.

\begin{figure}
\begin{centering}
\includegraphics[width=\columnwidth]{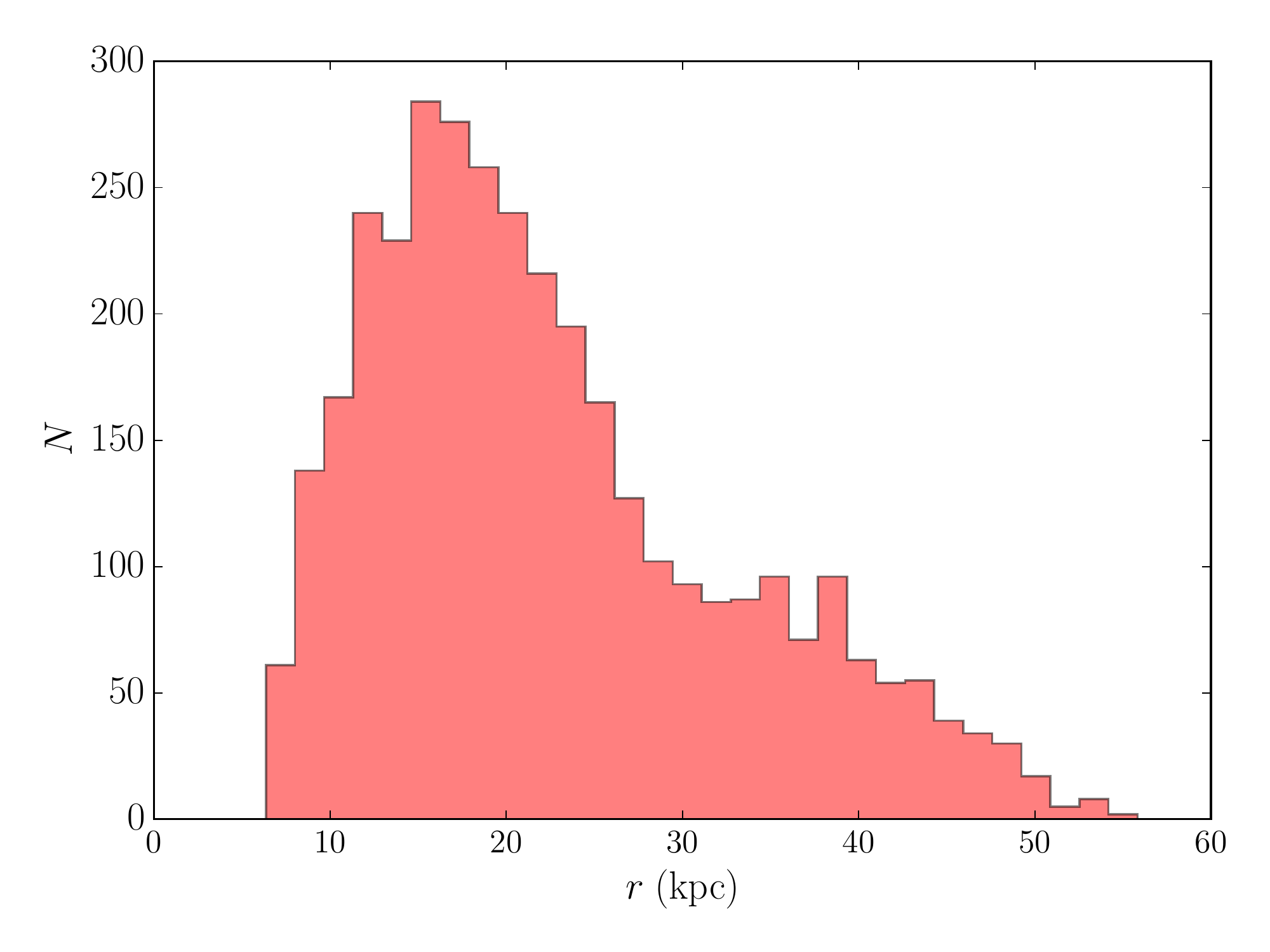}\newline\includegraphics[width=\columnwidth]{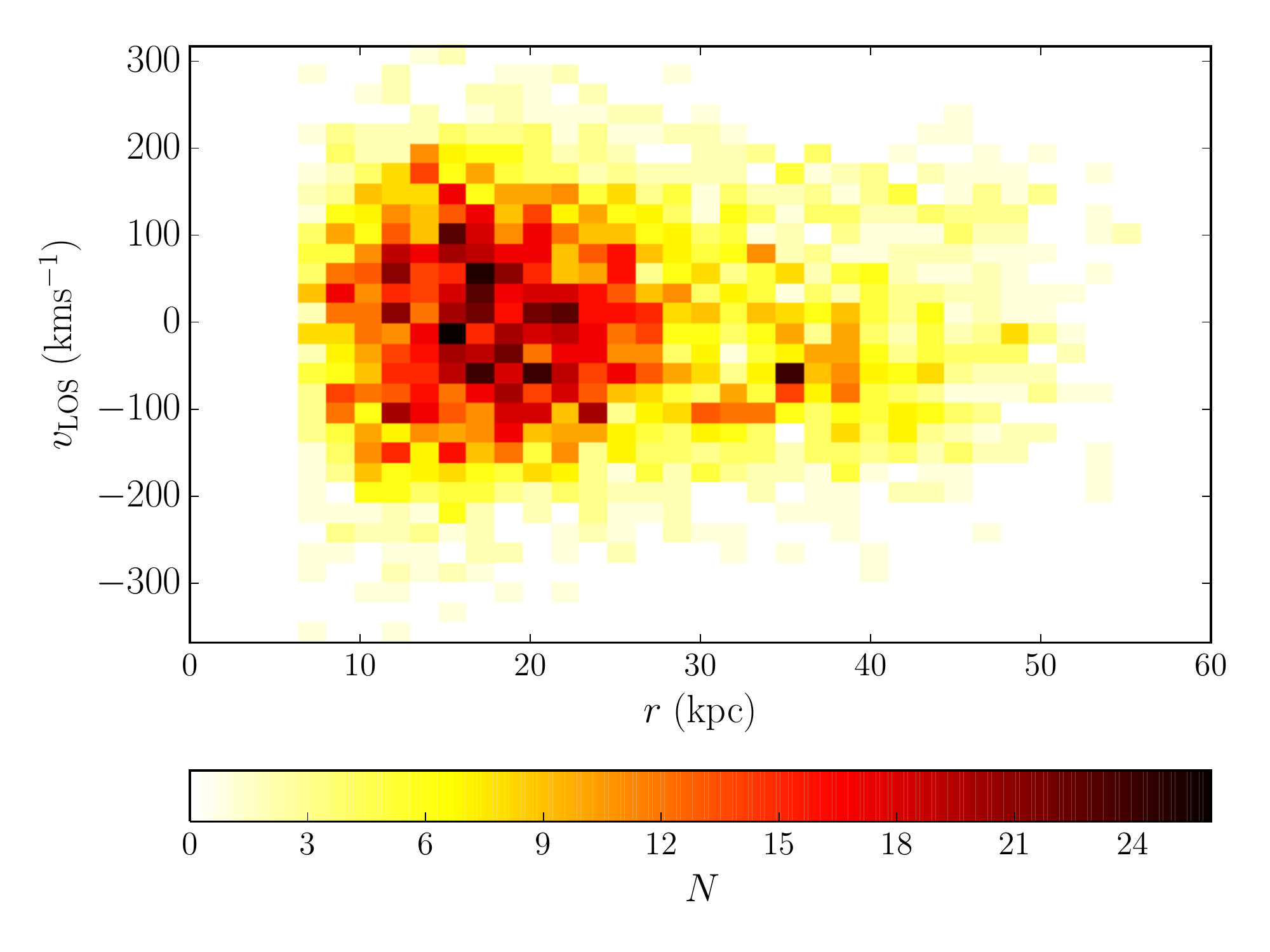}\newline\includegraphics[width=\columnwidth]{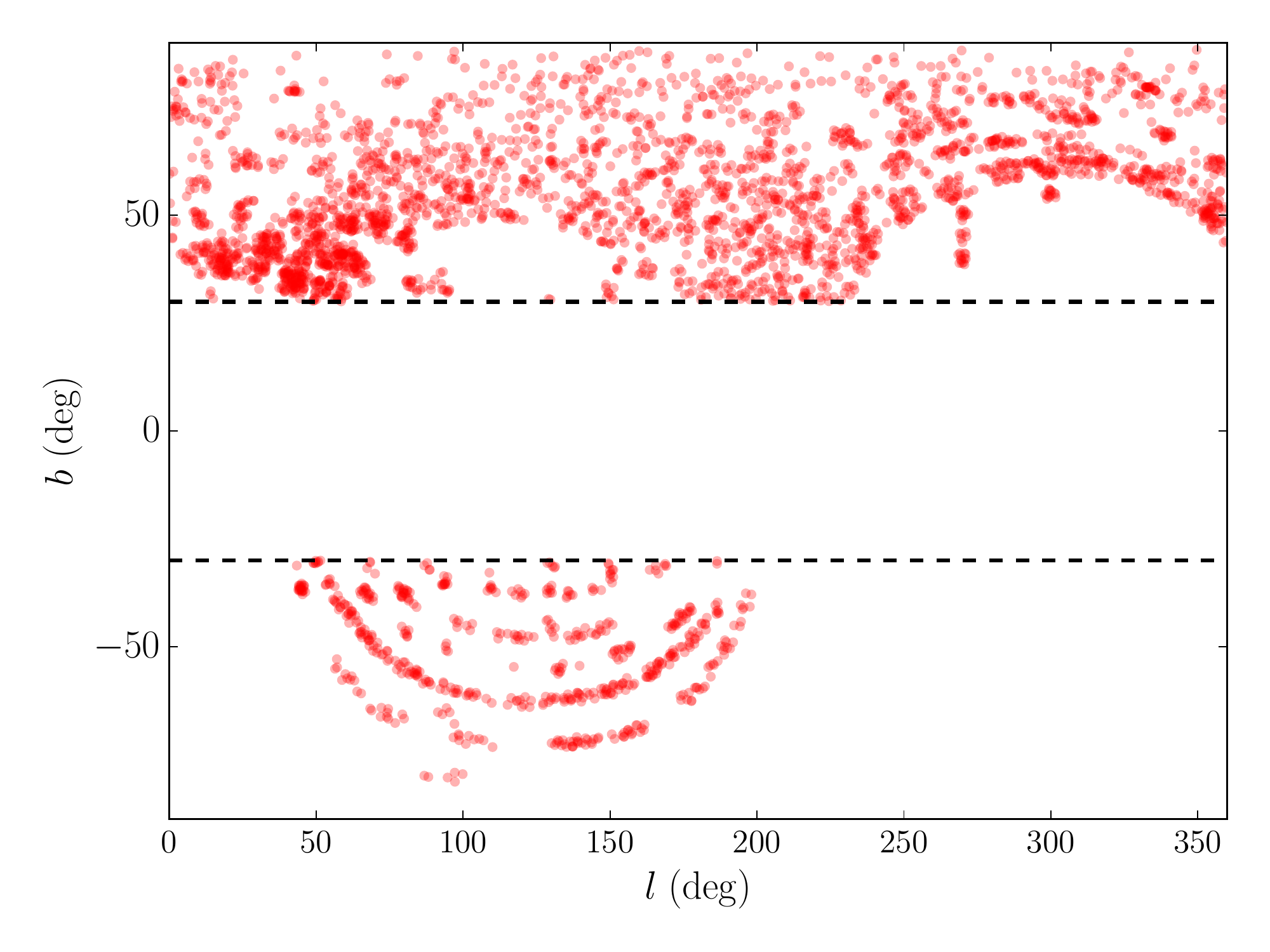}
\caption{Distributions of stars in the BHB dataset. Top: Histogram of
  Galactocentric distance. Middle: Distribution of Galactocentric
  distance and line--of--sight velocity. Note there is a noticeable
  overdensity at $r \sim 35 \kpc$ due to Sagittarius stream members.
  Bottom: Positions on the sky of the BHB sample after our cuts have
  been made. The dashed horizontal lines are at $|b|= 30 \degree$,
  which define our cut in Galactic latitude in order to avoid disc
  stars.}
\label{fig:sample_properties}
\end{centering}
\end{figure}

\subsection{Distribution function and gravitational potential}

Given the observational constraints, we wish to construct a DF that
generates a flattened density profile with a rapidly changing logarithmic slope, and
with a velocity ellipsoid that is is spherically aligned and radially
biased. Our model consists of a DF for the BHB population and a
parameterisation of the Galactic gravitational potential. We choose to
model the potential in a very simple way, using a spherical power law
\begin{equation}
\Phi (r) = -\dfrac{v_0^2}{\delta} \left(\dfrac{r}{10 \, \kpc}\right)^{-\delta},
\label{eq:galpot}
\end{equation}
where $0 < \delta < 1$. Although the true Galactic potential is
significantly more complex, this simple approximation is motivated by
recent studies of the Galactic potential far from the disk. For
example, in a recent study of the GD1 stream, \citet{Bo15} found the
potential to be well described by a scale--free and modestly flattened
($q = 0.91 \pm 0.04$) model at distances $r_\mathrm{GC} \sim 15
\kpc$. Using the same stream generation method, \citet{Gi14} showed
that the mass profiles of flattened model galaxies are nonetheless
accurately recovered using a model potential that is
spherical. Furthermore, the spherical alignment of the velocity
ellipsoid is also suggestive that the gravitational potential is
itself close to spherically symmetric at large radii (Evans et
al. 2015). That said, our over--simplification of the Galactic
potential is only permissible in a first attempt to fit data with
action-based DFs, and should be superseded in later work. In particular, 
evidence on the shape of the dark matter halo of the Galaxy is 
confused, and trivial as well as axisymmetric shapes remain 
possible~\citep[e.g.,][]{La09, La10, Ve13, De13}.

In a spherical potential, the actions are the azimuthal action
$J_\phi$, which is equal to the angular momentum about the $z$-axis;
the latitudinal action $J_\theta = L - |J_\phi|$, where $L$ is the
specific angular momentum, and the radial action
\begin{equation}
J_r = \dfrac{1}{2\rmpi}\int \sqrt{2\left[E-\Phi (r)\right] - L^2/r^2}\intd r.
\end{equation}
Since our choice of gravitational potential is a power-law, the result
found by \citet{Wi14a} applies, which gives the Hamiltonian of a
generic power law as a function of the actions
\begin{equation}
H(\acts) = C\left(L + DJ_r\right)^\zeta = C\mathcal{L}^\zeta,
\end{equation}
where $C$, $D$ and $\zeta$ are functions of the power law index
$\delta$ (for full expressions see \citealt{Wi14a}). \citet{Wi15} then
showed that a model with two power--law regimes is generated by a DF
of the form
\begin{equation}
f(\acts) \propto \dfrac{\mathcal{L}^{-\lambda}}{\left(\mathcal{L}^2 + \JB^2\right)^{(\mu - \lambda)/2}},
\label{eq:WEdf}
\end{equation}
where $\JB$ is the `break action', and controls the location of the
break in the density profile of the model. We expect our DF to broadly
resemble this ansatz, although here we are modelling a tracer
population in an independent potential rather than building a
self--consistent model. The DF of Equation (\ref{eq:WEdf}) generates a
spherically symmetric, isotropic double power-law model in the
potential of Equation (\ref{eq:galpot}), since $\mathcal{L}$ is very
nearly a function of energy alone. The parameters $\mu$ and $\nu$ are
related to the logarithmic slopes of the stellar halo density profile $\a0$ and
$\ainf$ by
\begin{eqnarray}
\lambda = \zeta\left(\a0/\delta - 3/2\right), \nonumber \\
\mu = \zeta\left(\ainf/\delta - 3/2\right).
\end{eqnarray}
In order to give the model an anisotropic, spherically aligned
velocity ellipsoid, we follow \citet{Wi15}, who demonstrated that
modifying the argument $\mathcal{L}$ to the DF such that
\begin{equation}
\mathcal{L} \rightarrow L + fDJ_r,
\end{equation}
($f>0$) endows the model with anisotropic kinematics. This arises
because the factor $f$ alters the relative importance of high and low
eccentricity orbits. If $0<f<1$, then the model becomes radially biased,
and $f>1$ gives tangentially biased models.

Our final modifications to the DF will have the effect of breaking the 
spherical symmetry of the model density profile, so that it becomes 
flattened. In action space, this means
introducing an explicit dependence on $J_\phi$, which will give the
model a symmetry axis. \citet{Br96} discovered a very elegant method
for this purpose, though only for scale--free models. Remarkably, they
found that introducing a multiplicative factor has precisely this
effect. This factor is given by
\begin{equation}
h(e^2 J_\phi^2/ L^2) = \sum\limits_{k=0}^\infty \dfrac{(1)_k (\frac{\alpha}{2})_k}{k! (\frac{1}{2})_k} \left(\dfrac{e^2 J_\phi^2}{L^2}\right)^k,
\label{eq:flatfactor}
\end{equation}   
where $e = \sqrt{1-q^2}$ is the ellipticity, $(...)_k$ is the
Pochammer symbol and $\alpha$ is the power law slope in the
density. This factor has the effect of reducing the number of stars on  
orbits where the ratio $J_\phi/(|J_\phi|+J_\theta)$ is small, which reduces the amount 
of vertical motion in the model relative to the circulation around the symmetry axis, thereby flattening 
the density profile. The expression at first sight appears unwieldy due to the
sum to infinity, but numerical experiments demonstrate that the series
converges very quickly, after $\sim 10$ terms. As it stands, this
expression is inappropriate for our purposes, since it is designed for
scale--free densities. We make the replacement
\begin{equation}
\alpha \rightarrow \alpha(\acts) = \dfrac{\a0 + \ainf (|\acts|/\JB)^2 }{1 + (|\acts|/\JB)^2},
\label{eq:varpow}
\end{equation}
where $|\acts| = \sqrt{\sum\nolimits J_i ^2}$. This approximately
preserves the constant flattening of the model. A suitable DF is
therefore
\begin{equation}
f(\acts) = \dfrac{\mathcal{N}}{\JB^{3-\mu}(2\rmpi)^3} \dfrac{h(\acts)\,\mathcal{L}^{-\lambda}}{\left(\mathcal{L}^2 + \JB^2\right)^{(\mu - \lambda)/2}},
\label{eq:ourDF}
\end{equation}
where $\mathcal{N}$ is a normalisation factor such that 
\begin{equation}
(2\rmpi)^3\int \,f(\acts)\,\intd ^3 \acts = 1, 
\end{equation}
and $h(\acts)$ is given by Equations (\ref{eq:flatfactor}) and
(\ref{eq:varpow}). We have now constructed a model with all of the
desired properties motivated by observations. All of our parameters
are nicely physical, except for the break action. In practise, we
should like to replace this parameter with a break radius,
$\rB$. Unfortunately, a wholly analytical expression for $\JB$ in
terms of a break radius is unavailable. We can approximate $\JB$ with
the expression
%
%\begin{equation}
%\JB \simeq e^{-1}J_1 + \left(1 - e^{-1}\right)J_\mathcal{-\infty},
%\end{equation}
% 
%where $J_1$ is the radial action of an orbit with no angular momentum
%and apocenter at $\rB$, and $J_\mathcal{-\infty}$ is the angular
%momentum of a circular orbit at $\rB$. When the model is completely
%built from circular orbits $\JB = J_\mathrm{-\infty}$, whereas in the
%purely radial case $\JB = J_1$. Since $J_1$ and $J_\mathcal{-\infty}$
%are analytically available in our model, we can write down a formula
%for the break action in terms of a scale--length $r_0$
%
\begin{equation}
\dfrac{\JB}{1000\kms\kpc} = \left(\dfrac{v_0}{100\kms}\right) \, \dfrac{\delta^{1/\delta}}{\sqrt{\delta}} \, \left[\zeta - \frac{\zeta}{2} \delta (1 - \mathrm{e}^{-1})\right]^{1/\zeta} \, \left(\dfrac{r_0}{10\kpc}\right)^{-\delta/\zeta}. 
\end{equation}
This expression is derived by considering the relative contributions of perfectly radial and circular orbits to the density profile of the model. For a given $r_0$ and $v_0$, the model density profile 
will break at an elliptical radius $\rB \simeq r_0$ if $\JB$ from the above expression is used as the break action. We then choose to sample the parameter $r_0$. The model parameters are then
\begin{equation}
\mathcal{P} = \left(\a0,\,\ainf,\,r_0,\,q,\,f,\,v_0,\,\delta\right),
\end{equation}
each of which has a simple physical interpretation.

\begin{figure}
\includegraphics[width=\columnwidth]{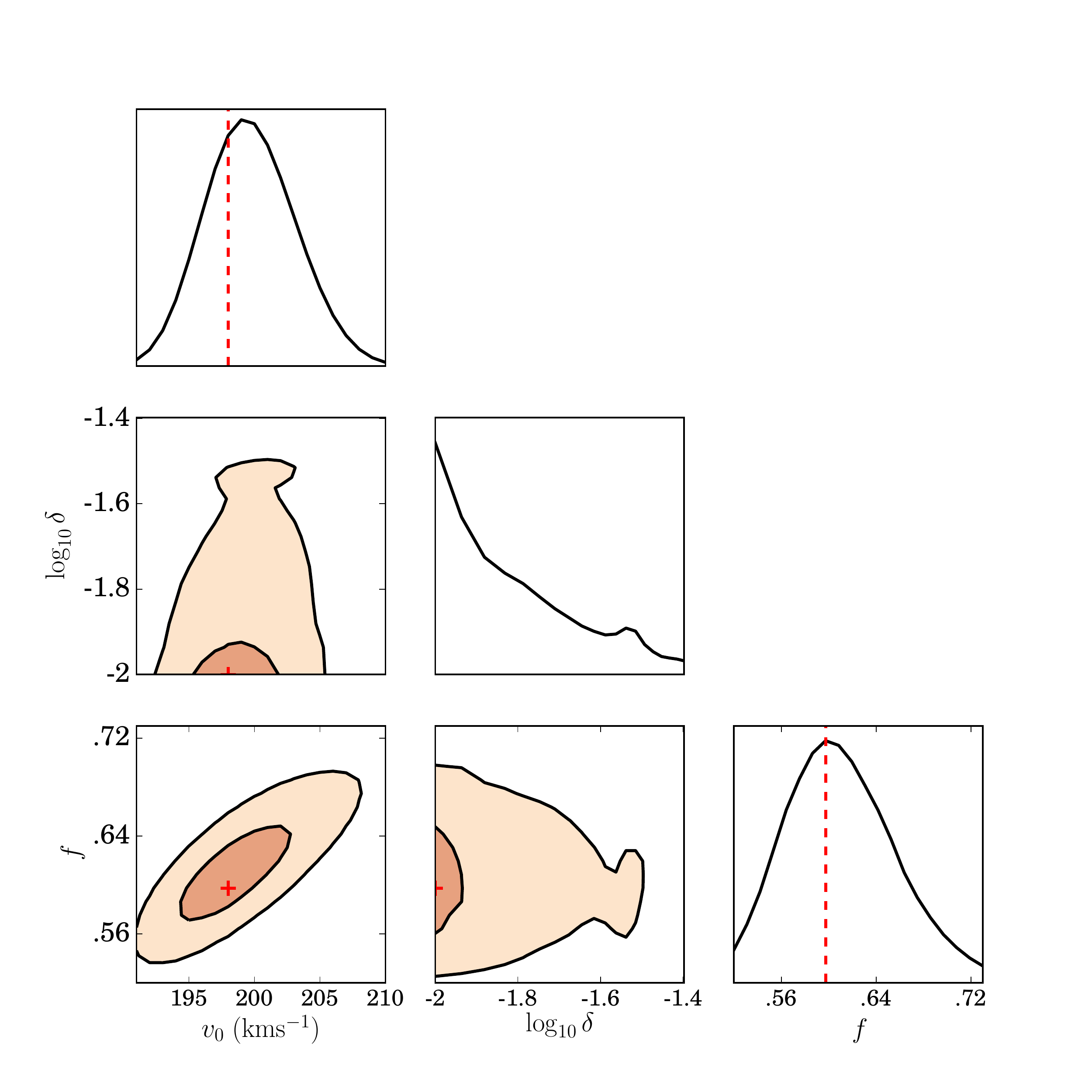}
\caption{Likelihood contours from our analysis. We
  show the 68\% and 95\% confidence intervals of the marginalised distribution
  over pairs of model parameters, and mark the maximum likelihood
  gridpoint with a red cross. In the one--dimensional distributions, the 
  maximum likelihood value for each parameter is marked with a dashed vertical line. 
  One can see that the circular velocity at $10\kpc$ is
  tightly constrained, and that models close to logarithmic potentials
  are favoured. The contours of marginalised distributions involving
  $\delta$ are one--sided due to the fact that we only consider
  declining rotation curves. Our grid in $\delta$ was truncted at 
  $\delta = 10^{-2}$ because of the highly singular limit that occurs at 
  $\delta = 0$.}
\label{fig:likelihoodconts}
\end{figure}

\begin{figure*}
\includegraphics[width=\columnwidth]{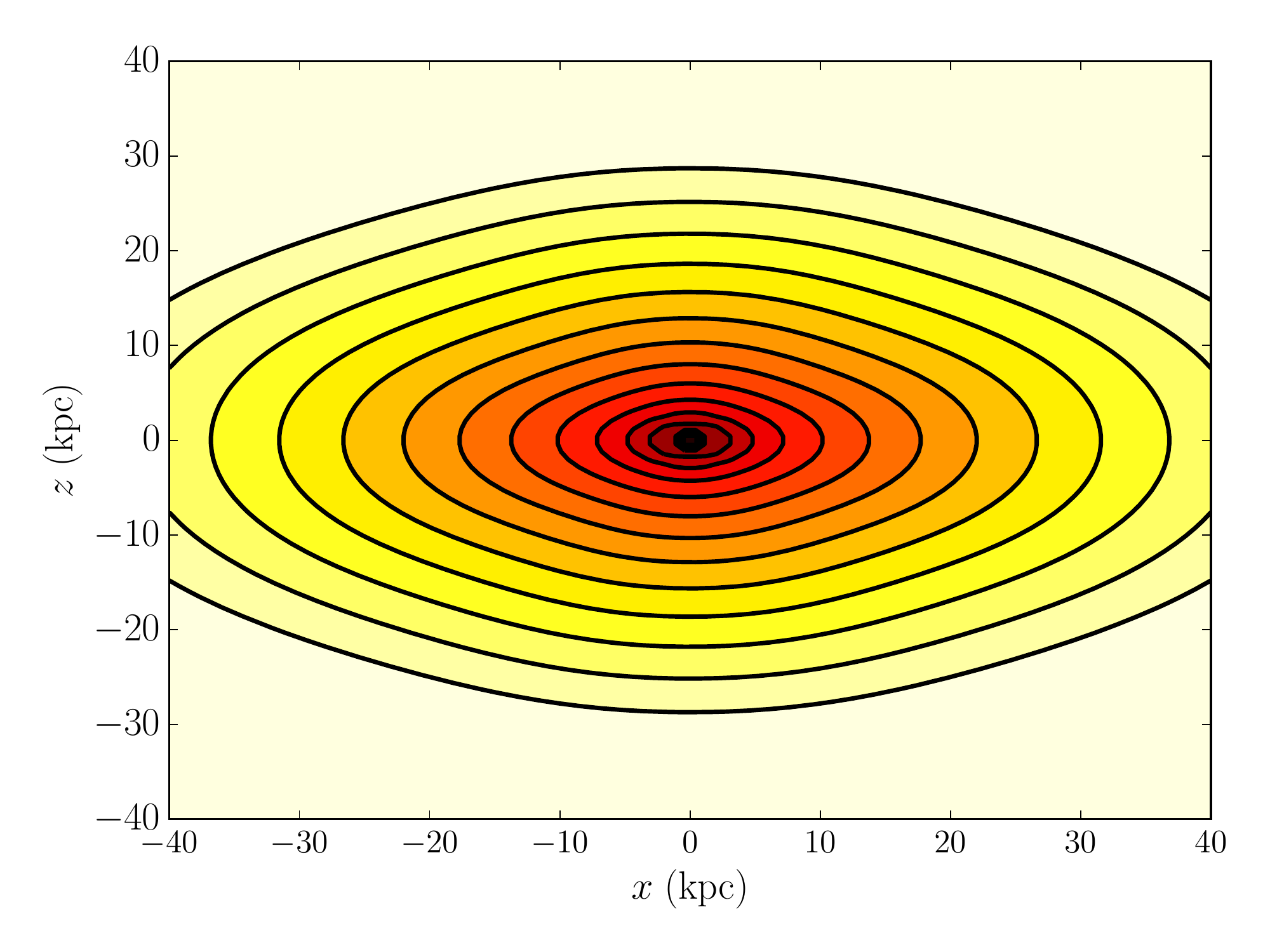}\quad\includegraphics[width=\columnwidth]{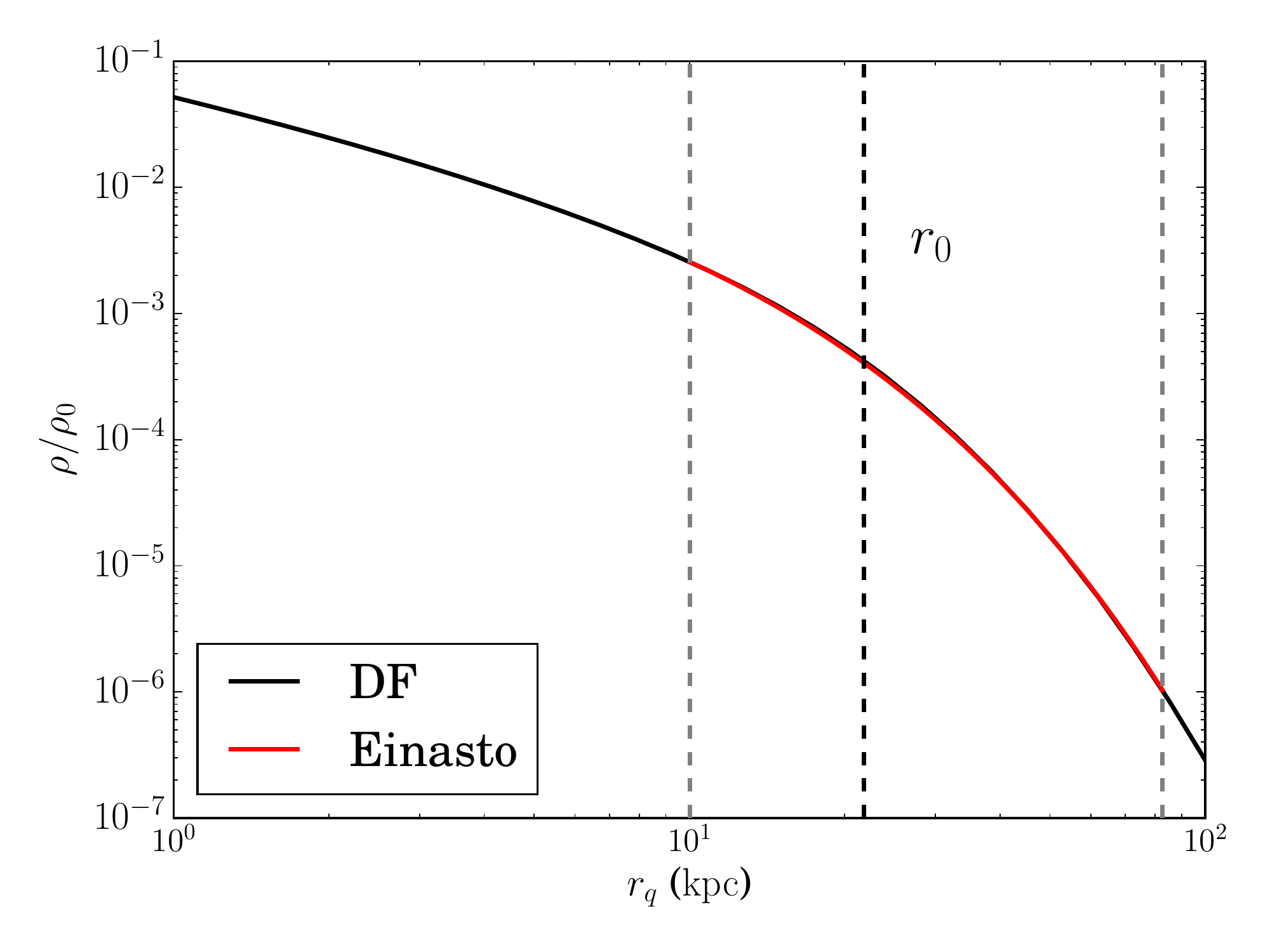}
\caption{Spatial properties of the model. Left: the density profile of
  the model in the $x$--$z$ plane. Right: the 1D density profile of
  the model with elliptical radius. The black dashed line denotes the
  value of the parameter $r_0$, and the grey lines surround the region
  for which we have data. Overplotted in the range where the data lies
  is the best--fit Einasto profile from \citet{De11c}, which we use to
  constrain the density profile of our model.}
\label{fig:spatialprops}
\end{figure*}

\section{Data}

\begin{figure}
\includegraphics[width=\columnwidth]{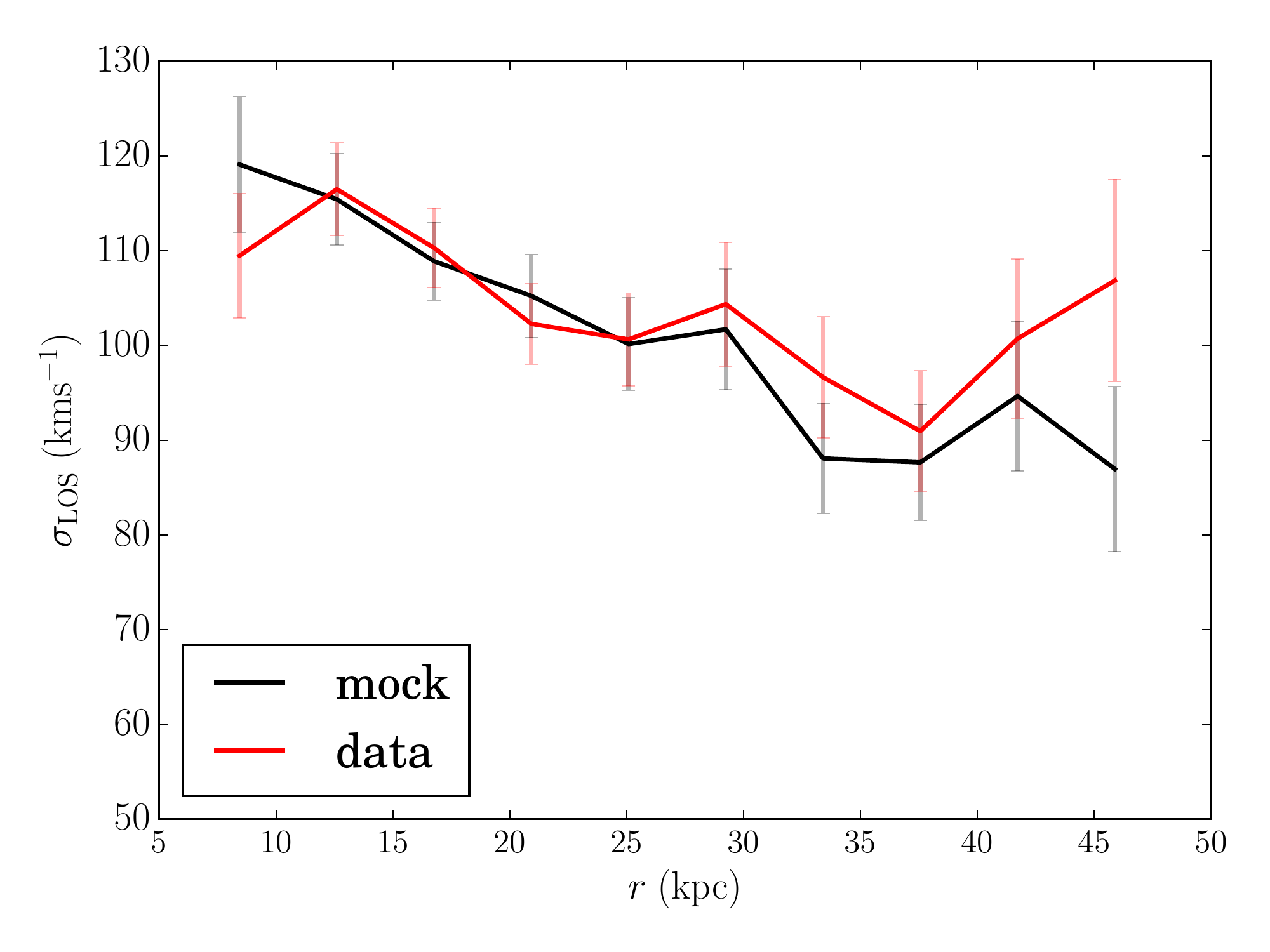}\newline\includegraphics[width=\columnwidth]{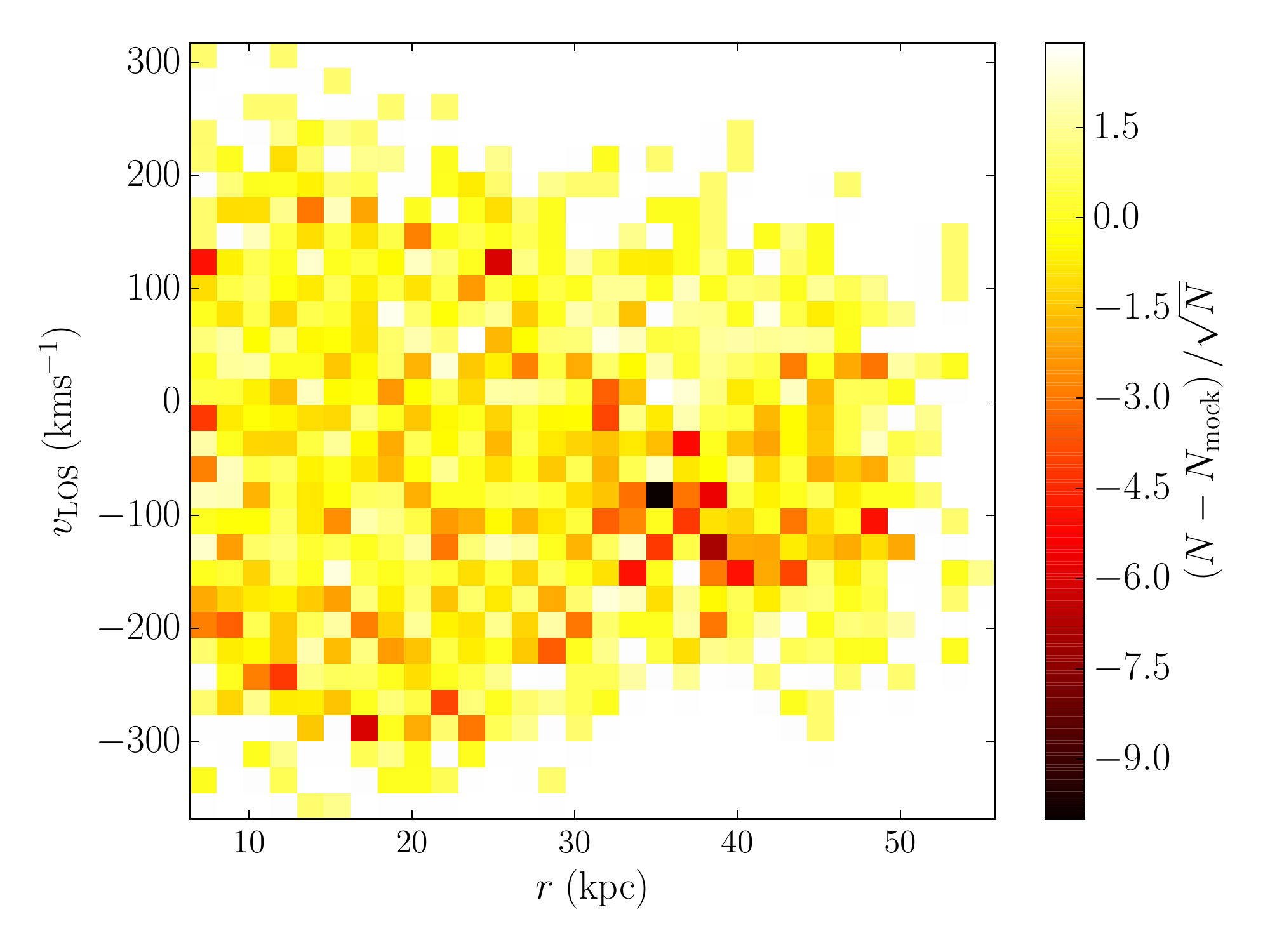}\newline\includegraphics[width=\columnwidth]{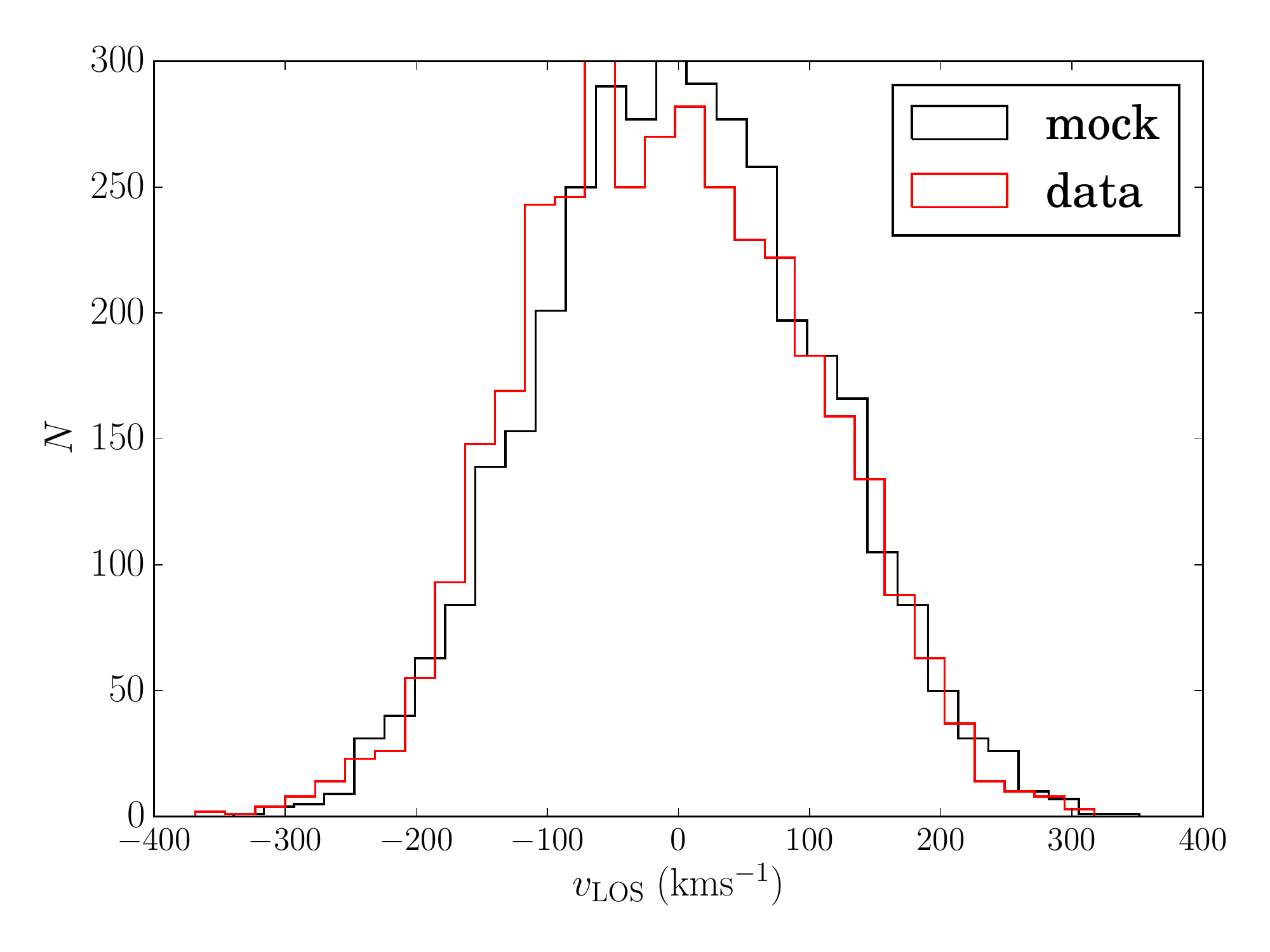}
\caption{Comparisons of our maximum--likelihood model to the data. Top
  panel: the line--of--sight velocity dispersions of the data and mock
  catalogue as a function of Galactocentric radius. The two profiles
  are in excellent agreement, other than an upturn in the profile of
  the data at large radii. Middle: residuals between the model and the
  data in the dsitribution of Galactocentric radius and
  line--of--sight velocity.  The residuals are essentially noise other
  than a feature between 30 and $50\kpc$ at $\sim -100\kms$, which is
  due to stars belonging to the leading arm of the Sagittarius
  stream. Bottom: comparison of the distributions in line--of--sight
  velocity between the model and the data. The two distributions are
  in good agreement, except the data displays a slight skew towards
  negative line--of--sight velocities, due to the overdensity of
  Sagittatius stars at $\vlos \sim -100\kms$.}
\label{fig:datamock}
\end{figure}

In this study, we fit our model to the dataset provided by
\citet{Xu11}, who compiled a catalogue of BHB
stars from the SDSS DR8 catalogue \citep{Ai11}. The dataset contains
$\sim 4000$ spectroscopically confirmed BHB candidates, arising from
various spectroscopic surveys within SDSS. BHB stars are a very useful
class of object to use when investigating the stellar halo, as they
are intrinsically bright, with $M_g \sim 0.5$, and \citet{De11c}
showed that very accurate estimates of their absolute magnitudes can
be derived ($\Delta M_g \sim 0.15$) via the polynomial relation
\begin{eqnarray}
M_g &=& 0.434 - 0.169(g-r) + 2.319(g-r)^2 \nonumber \\
	&+& 20.449(g-r)^3 + 94.517(g-r)^4,
\label{eq:absmag}
\end{eqnarray}
which is valid in the colour range $-0.25 < g-r < 0.0$. From the
absolute magnitude, we can estimate distances to the stars in our
sample to high precision. Before carrying out our analysis, we make
two cuts on the dataset. First, we limit the colours to the range for
which Equation (\ref{eq:absmag}) is valid, to ensure that the inferred
distances are reliable. We then only include stars for which $|b| >
30^\circ$ in order to reduce contamination from the disk. This leaves
us with 3534 stars at Galactocentric radii $10 \kpc \lesssim r
\lesssim 50 \kpc$, each of which has 4 of 6 phase--space coordinates
very accurately constrained. Figure \ref{fig:sample_properties}
depicts the distributions in Galactocentric distance, line--of--sight
velocity and on--sky positions of the data.

\section{Fitting Methodology}

Ideally, we would like to use this dataset to constrain all the
parameters of our model. This would mean inferring properties of the
entire phase--space distribution of the stellar halo, including the
density profile and kinematics. However, in order to carry out such a
study, we would require detailed knowledge of the selection function
of the sample of stars that is being used. The observed distribution
of stars in phase--space and chemistry, $\boldsymbol{Z}$, is the
product of the selection function with the true density, i.e.
\begin{equation}
f_\mathrm{obs}\left(\boldsymbol{x},\,\boldsymbol{v},\boldsymbol{Z}\right) = S\left(m\,,\,l\,,\,b\,,\,\boldsymbol{Z}\right)
																			\times f_\mathrm{true}\left(\boldsymbol{x},\,\boldsymbol{v}, \boldsymbol{Z} \right).
\end{equation}
Note that the selection function, $S$, is presumed not to depend on
the velocities of the stars, but does depend on their magnitudes ($m$),
on--sky positions and chemistry. Without an effective model of $S$, we
cannot hope to constrain the spatial and chemical distributions of the
stars. Unfortunately, the selection function for our data is
essentially unknown. Since the stars were observed on a variety of
different plates from different spectroscopic surveys, and they have a
relatively low on--sky density, we concluded that a
reverse--engineered selection function of the sort used by
\citet{Bo12} cannot be produced. This means inference on the density
profile of the stellar halo is not possible with the data used here.

All hope is not lost, however. Although some of the parameters of our
model cannot be constrained by this data, we can still proceed. We can
use the independent analysis of \citet{De11c} to place strong priors
on the spatial distribution of the stars, and then proceed to use the
velocities of the stars to constrain the kinematics of the stellar
halo, as well as the Galactic mass profile.  This is precisely what
was done by \citet[][hereafter D12]{De12}, though they used a smaller
subsample of $\sim 1900$ stars from this dataset.  Our DF lets us
model the entire dataset, and thus place tighter constraints on halo
kinematics and the mass of the Galaxy, though of course the properties
of the stellar halo such as the flattening ($q= 0.59$) are fixed at
outset.

Given the above discussion, we use the data to constrain only 3 of the
7 model parameters: $f$, which controls the kinematics, $v_0$, the
circular speed at $10\kpc$, and $\delta$, the slope of the
gravitational potential. The remaining parameters are fixed for each
set of $(f\,,\,v_0\,,\,\delta)$ by optimizing the fit of our model's
density profile to the Einasto model of \citet{De11c} described in
Section \ref{sec:obsmot}. This stage of the procedure is equivalent to
enforcing a very strong prior on the remaining 4 parameters of the
model. The likelihood for an individual datum is then given by
\begin{eqnarray}
p(v_\mathrm{LOS,\,i} \, | \, \boldsymbol{x}_i \, , \, f \, , \, v_0 \,
, \, \delta) &=& \int f(\acts) \left| \dfrac{\partial (\acts\,,\,
  \boldsymbol{\theta})}{\partial (\boldsymbol{X})}\right| \, \intd v_l
\, \intd v_b \nonumber \\ &/& \int f(\acts) \left| \dfrac{\partial
  (\acts\,,\, \boldsymbol{\theta})}{\partial (\boldsymbol{X})}\right|
\, \intd v_l \, \intd v_b \intd v_\mathrm{LOS},
\end{eqnarray}
where $\boldsymbol{X}$ are the usual Galactic coordinates. The above
expression simplifies, owing to the fact that the Jacobian factors
depend only on position coordinates, giving
\begin{equation}
p(v_\mathrm{LOS,\,i} \, | \, \boldsymbol{x}_i \, , \, f \, , \, v_0 \, , \, \delta) = \dfrac{1}{\rho(\boldsymbol{x}_i)} \, \int f(\acts) \, \intd v_l \, \intd v_b,
\end{equation}
which is the normalised line--of--sight velocity distribution at the
position of the star in question. The log--likelihood for the entire
dataset is then simply
\begin{equation}
\log l  = \sum\limits_{j=1}^{N_\mathrm{BHB}} \log p_j. 
\end{equation}
In order to compute this likelihood with minimal noise,
high--precision numerics are required for a number of different
tasks. First, we need to evaluate the non--classical integral $J_r$
in order to compute the integrand $f(\acts)$. In order to do this, we
compute $J_r$ numerically using adaptive Gaussian quadrature
implemented in the {\sc Gnu} scientific library over a grid of 200
points in angular momentum at some fiducial energy $E_0$. Since the
potential is scale--free, the radial action scales self--similarly
with energy, so we do not need to consider other energies: $J_r$ can
be recovered by simple multiplication by a scale factor. In--between
grid points, we recover $J_r$ using cubic spline interpolation.

Once the radial action is available, we must perform the two
(marginalising over the proper motions) and three--dimensional
(integrating over velocities to find the density) integrals required
to compute the normalised line--of--sight velocity distributions. This
is done using the Cuhre algorithm implemented in the {\sc Cuba}
package \citep{cuba}. This particular algorithm is deterministic,
which guarantees that our likelihoods are replicable.

Evaluating the triple integral needed to calculate the density 
of the model at a given position is significantly more time-consuming than 
the double integral when marginalising over the proper motions. However, the 
model has a smooth two--dimensional density profile, and we can use this to 
our advantage. Instead of explicitly evaluating the density at the positions of 
all 3534 stars, we instead construct a grid and use bilinear interpolation to 
recover the density. An astute choice of variables proves to be
\begin{eqnarray}
u &=& \frac{1}{2}\log \left( R^2 + (z/q)^2 \right), \\
t &=& \arctan \left( \dfrac{qR}{|z|} \right). \nonumber
\end{eqnarray}
An economically sized grid can be used for interpolation to find the density. We use 
a grid of 20 points in $u$ and 10 points in $t$, and find that the density at the 
position of any star is recovered to a precision of 0.4 percent at worst.

All of the above calculations are performed in {\sc C}, which is then
wrapped for use in {\sc python}. In order to locate the
maximum--likelihood solution, we perform a grid--search. Monte--Carlo
methods are unnecessary in this instance, due to the low
dimensionality of the parameter space.

\section{Results \& Discussion}

\begin{table*}
\begin{tabular}{ccccccc}
\hline

$\a0$ & $\ainf$ & $r_0$ & $q$ & $f$ & $v_0$ & $\delta$ \tabularnewline
\hline
$0.84^{+0.07}_{-0.30}$ & $9.13^{+0.96}_{-0.19}$ & $21.6^{+1.5}_{-1.5}\kpc$ & 0.59 & $0.59^{+0.07}_{-0.04}$ & $198.2^{+3.4}_{-3.2}\kms$ & $0.01^{+0.006}_{-0.01}$ \tabularnewline
\hline
\end{tabular}\caption{Parameters of our maximum likelihood model and their dispersions. We do not quote a dispersion for the flattening, $q$, because it is kept fixed 
during the fitting procedure.}
\label{table:parameters}
\end{table*}

Here, we present the results of our analysis. First, we consider how
well the model fits the data. Then we discuss the constraints placed
on the Milky Way mass distribution and circular velocity profile,
comparing our results with those from other analyses. finally, we
discuss the kinematics of our best--fit model.

\subsection{Comparison to the data}

Figure \ref{fig:likelihoodconts} depicts the likelihood distributions from our analysis, and
Table \ref{table:parameters} summarises the best--fit parameters and
their errors. Note that four of the parameters are fixed by our prior
on the density profile, and Figure \ref{fig:spatialprops} depicts the
one and two--dimensional properties of the density of this model.

In order to compare the model to the data, we generated a mock catalogue 
by drawing a line--of--sight velocity at the position in the Galaxy of each star 
from the maximum likelihood model. The reason we do not also 
sample position is because, as previously stated, the selection function is 
unknown and depends on the location of the star. We have assumed that the 
selection function does not depend on the kinematics of a star, and we are 
therefore safe to freely sample the line--of--sight velocity distributions 
of the model at positions where stars have been observed. We draw the mock catalogue 
using a simple rejection sampling technique.

Figure \ref{fig:datamock} depicts several comparisons between the data
and our mock catalogue.  The upper panel, depicting the
line--of--sight velocity dispersions as a function of galactocentric
radius, shows excellent agreement between our model and the
data. There is an upturn in the profile of the data at large radii,
which is due to the presence of stars belonging to the Sagittarius
stream. This substructure should then be apparent in the
two--dimensional distribution in galactocentric radius and
line--of--sight velocity. Indeed, as the middle panel of Figure
\ref{fig:datamock} demonstrates, there is a signal at the same
distances at $\vlos \sim -100 \kms$. The final panel depicts the
overall distribution in line--of--sight velocity. The profiles are in
good agreement, except the data shows a slight skew towards negative
line--of--sight velocities, which is again due to the presence of
Sagittarius members.

We confirmed that the systematic residuals between the data and the mock 
catalogue are a consequence of the Sagittarius stream by making a further cut 
to the dataset. We removed stars with $|B|<10^{\circ}$, where $B$ is one of the 
two Sagittarius stream coordinates $(\Lambda,B)$ defined in \citet{Be14}. First, we inspected 
the line--of--sight velocity distribution of the stars beyond $r = 30\kpc$, and found that the skew 
towards negative radial velocities was removed by this cut. We then 
drew another mock catalogue from the model and compared it to the reduced dataset. The signal seen 
at $\vlos \sim -100 \kms$ is no longer present, confirming that the contamination 
in the sample is largely from the Sagittarius stream.

Overall, the model seems to be in good agreement with the data, and we
are confident that the Sagittarius stream members do not bias our
fits, since the majority of our constraining power comes from stars at
galactocentric radii $< 30\kpc$, for which we do not possess
significant contamination. In fact, this highlights an interesting application of smooth models of the 
stellar halo: one can fit a smooth model to the data, and search for substructure using the residuals 
between the data and the model. Having established that our model is as good representation of the data, 
we can now go on to discuss the implications of our inference.

\subsection{The Cumulative Mass Profile of the Milky Way}

\begin{figure*}
\includegraphics[width=\columnwidth]{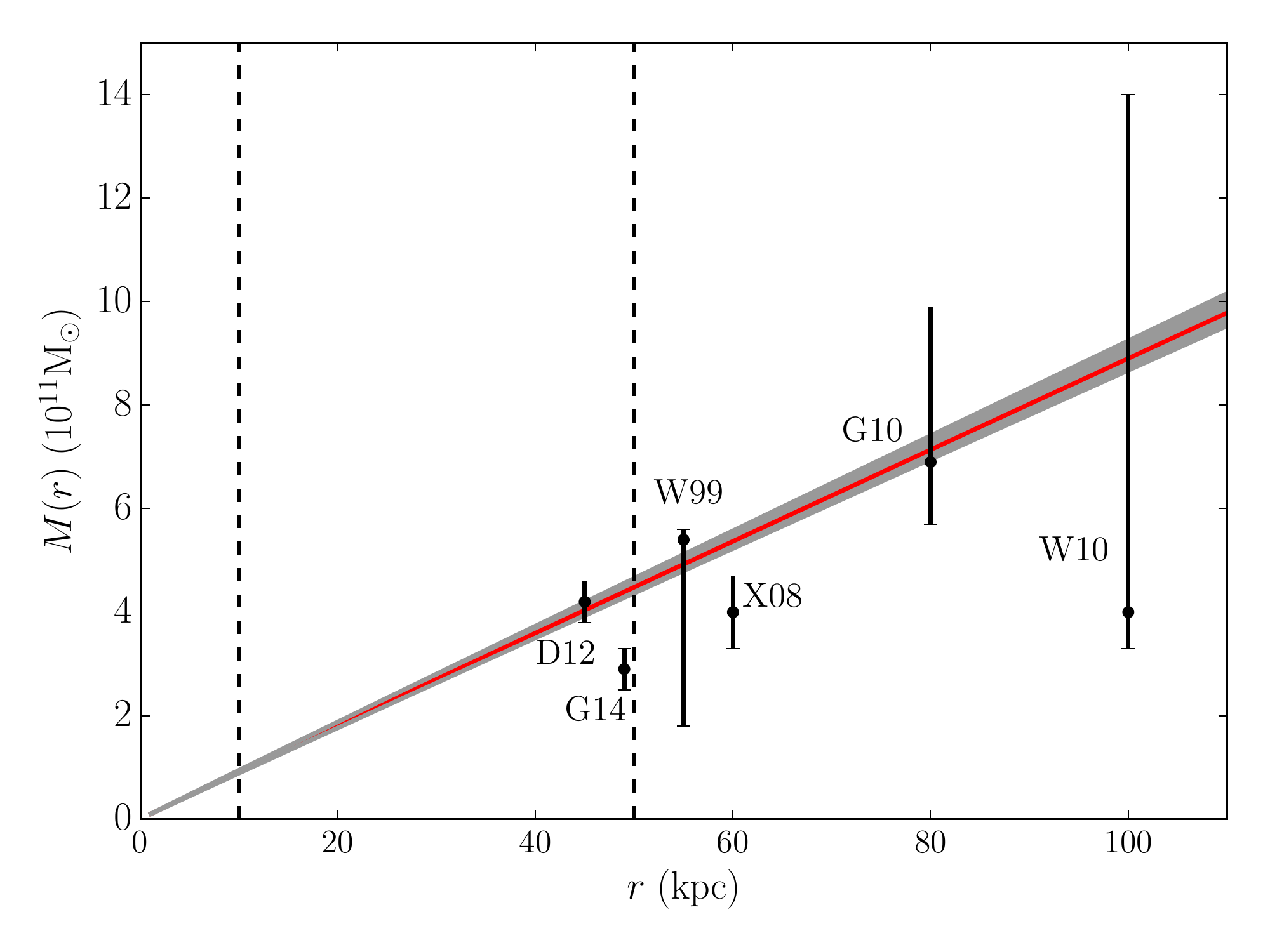}\includegraphics[width=\columnwidth]{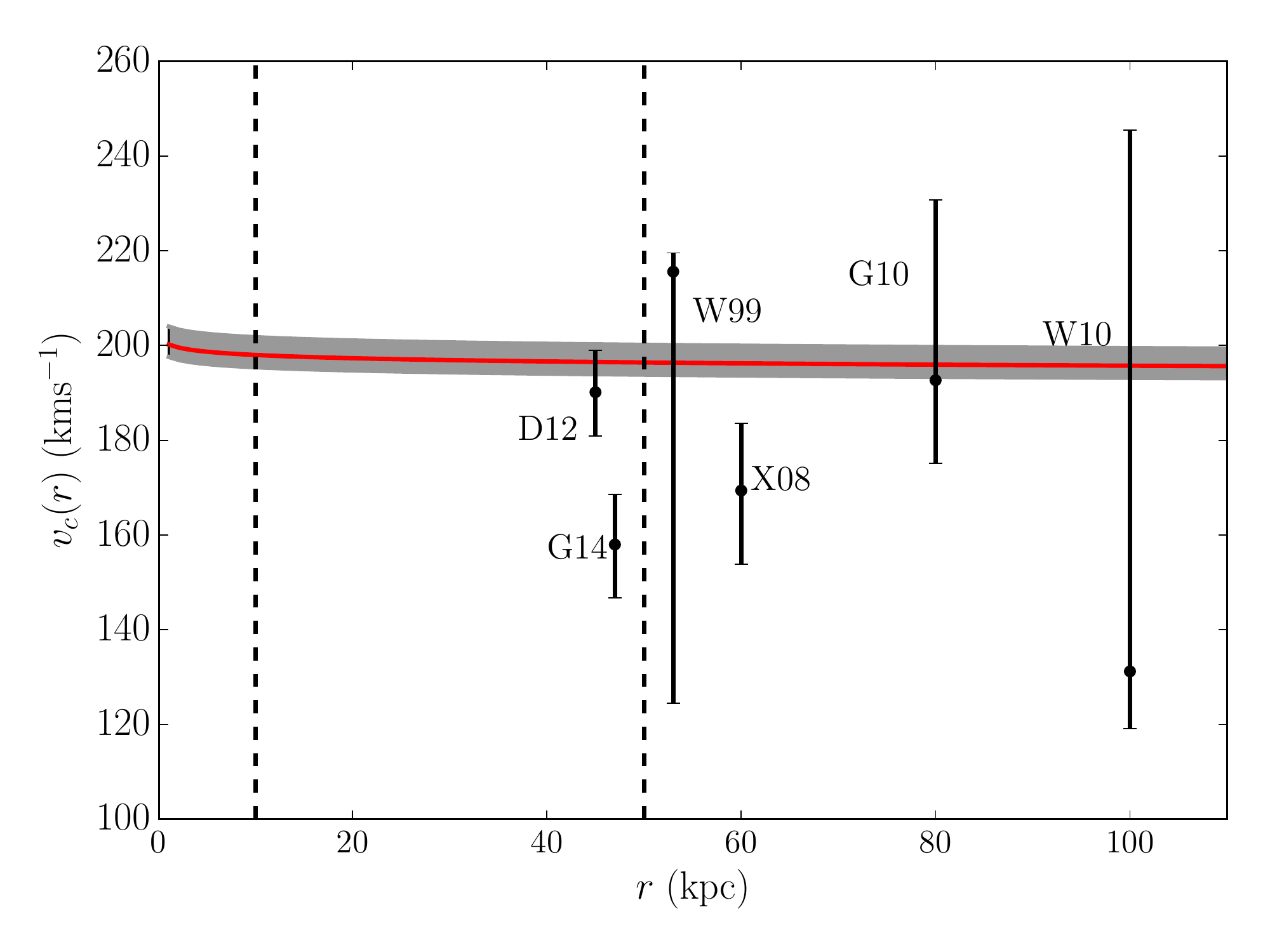}
\caption{Maximum likelihood profiles (red curves) and 68\% confidence
  intervals (grey shade) for the mass profile (left) and circular
  velocity curve (right) of the Milky Way. The error bars come from sampling the 1
  $\sigma$ confidence interval on the joint posterior of the two
  potential parameters. Dashed lines surround the
  region in Galactocentric radius for which we possess data. Also
  plotted are the quoted error bars from several other studies of the
  Milky Way cumulative mass distribution: D12 \citep{De12}, G14
  \citep{Gi14}, W99 \citep{Wi99}, X08 \citep{Xu08}, G10 \citep{Gn10}
  and W10 \citep{Wa10}. The markers corresponding to D12, G14 and W99
  are all offset from $50\kpc$ so that they are distinguishable.}
\label{fig:mass_vcirc}
\end{figure*}

 The parameters $v_0$, the spherically averaged circular speed at $10\kpc$,
and $\delta$, the slope of the potential, are tightly
constrained by our analysis. $v_0$ is found to be $\sim 200 \kms$, and $\delta$ is
very close to zero, implying the rotation curve is almost flat. Taking
these results together, we find that the predicted enclosed mass
within a spherical shell of radius $50\kpc$ is $4.48_{-0.14}^{+0.15}
\times 10^{11} M_\odot$.

Figure \ref{fig:mass_vcirc} depicts the cumulative mass distribution
and circular velocity curves of the Milky Way predicted by our model,
along with the 68\% confidence intervals. We have also
plotted the results from other studies on this matter, which we shall
now discuss in detail.  There are multiple predictions of the Milky
Way enclosed mass at $\sim 50\kpc$. D12 and \citet[][hereafter
  X08]{Xu08} both also use samples of BHB stars from the SDSS: D12
used a subsample of $\sim 1900$ stars from the same dataset considered
here, and X08 used a sample from an earlier data release. Our result
agrees extremely well with that of D12, who also used distribution
function based modelling to place constraints on the mass profile. As
one might expect, our error--bar is somewhat smaller, owing to the
fact that we use roughly double the amount of data in our
study. Interestingly, the logarithmic slope of the potential inferred
by D12 is larger than our result: $\delta \sim 0.4$. Their value is
almost half--way between the Keplerian and logarithmic cases, so the
rotation curve has a steeper decline with radius. This is an
interesting result, and has several possible explanations. It is clear
that the gravitational potential of the Galaxy cannot possibly be as
simple as is implied by Equation (\ref{eq:galpot}), and so biases
could certainly be introduced into an analysis that assumes such a
parameterisation. In particular, if the slope of the Galactic rotation
curve changes appreciably over the range of distances between 10 and
$50\kpc$, the lack of flexibility in the model means this cannot be
accounted for. D12 only considered stars beyond an elliptical radius
$r_q = 27\kpc$, the more distant stars in this dataset, so it is not
unreasonable to assume that their analysis was more sensitive to the
slope of the potential at larger distances. In our case, the
left--hand panel of Figure \ref{fig:sample_properties} informs us that
the majority of the stars reside at radii $10\kpc \lesssim r \lesssim
25\kpc$. Therefore, it may possibly be the case that the rotation
curve is approximately flat between 10 and $25\kpc$, but begins to
decline more sharply thereafter.

X08 took a rather different approach in their modelling of the BHB
population. Rather than fitting analytical distribution functions,
they instead compared their data to mock samples drawn from
cosmological simulations. Although they also predict a very flat
rotation curve, its amplitude is somewhat lower. This discrepancy
could be accounted for by the above reasoning, where the circular
velocity begins to decline more rapidly with radius.

\citet[][hereafter W99]{Wi99} estimated the mass enclosed at $50\kpc$
by modelling the distribution of the satellite galaxies and globular
clusters of the Milky Way. They used constant anisotropy tracer
distribution functions embedded in spherically symmetric models with
rotation curves that are flat in the inner parts, then decline in a
Keplerian fashion at large radii. These models have a circular
velocity of form:
\begin{equation}
v_{\rm c}^2 = \dfrac{v_0^2}{\sqrt{1+r^2/a^2}}.
\end{equation}
Their result, $M(r<50\kpc) = 5.4^{+0.2}_{-3.6} \times 10^{11}
M_\odot$, is in excellent agreement with ours (note the asymmetric
error--bars). It is also interesting to note that the scale--radius of
their model is found to be very large ($140\kpc \lesssim a \lesssim
260\kpc$), implying a very flat rotation curve at the distances of the
stars in our dataset.

\citet{Gi14} used observations of the Sagittarius stream, in
particular the precession of the apocentric position of the stream, to
infer the mass profile between radii $50\kpc < r < 100\kpc$. As the
title of their paper states, they find a particularly `skinny' Milky
Way, with an enclosed mass at $50\kpc$ of just $2.9 \pm 0.4 \times
10^{11} M_\odot$. Given the small error bars on their measurement,
their result is in tension with ours. Evidently, the two methods
possess different systematic biases.

We now compare results from two other studies, \citet[][hereafter
  G10]{Gn10} and \citet[][hereafter W10]{Wa10}, who both inferred the
mass enclosed at larger distances. Although it is still interesting to
discuss these results, our result must be extrapolated in order for
comparisons to be made. G10 used a large sample of radial velocity
measurements of halo stars to carry out a Jeans analysis. They assume
a spherically symmetric tracer density with logarithmic slope between
-3.5 and -4.5 with a constant anisotropy, and find $M(r<80\kpc) =
6.9^{+3.0}_{-1.2} \times 10^{11} M_\odot$. In spite of their less
sophisticated model for the tracer density, their result is in good
agreement with ours. Finally, \citet{Wa10} took a sample of 26 satellite
galaxies of the Milky Way and applied their virial mass estimators to
the data. The estimators assume a power--law tracer density, as well
as a power--law gravitational potential. Their measure is very
sensitive to the assumed anisotropy of the tracer population, and so
their estimate for the enclosed mass has a large error
bar. Nonetheless, we are consistent with their conclusions at the 68\%
confidence level.

\begin{figure}
\includegraphics[width=\columnwidth]{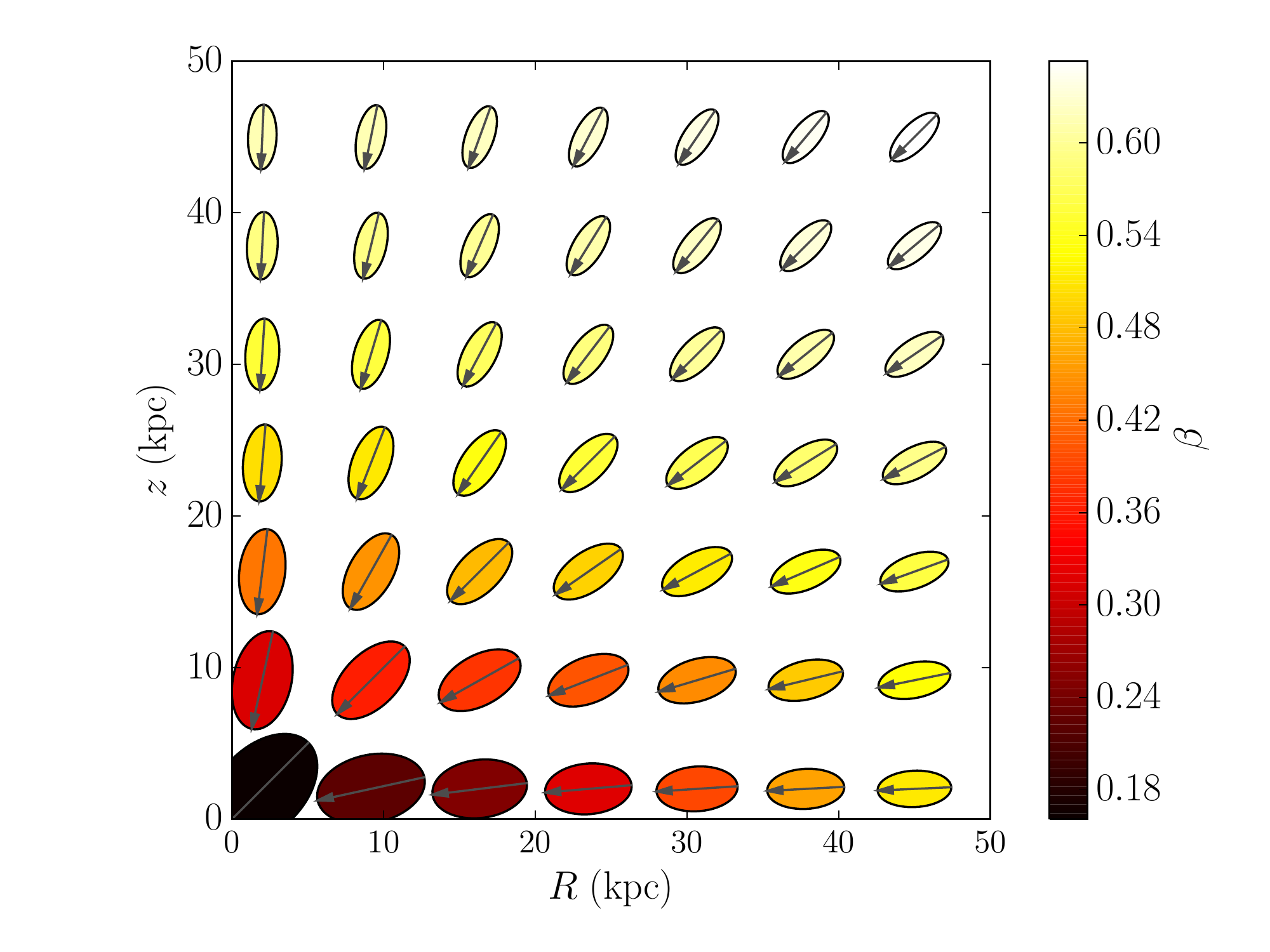}
\caption{Velocity ellipsoids of the best--fit model across a variety
  of positions. The ellipsoids are everywhere aligned with spherical
  coordinates, with long axes directed towards the Galactic
  centre. The size of the ellipsoids do not vary a great deal with
  radius, suggesting a relatively flat velocity dispersion
  profile. Their shape becomes more elongated with radius, indicating
  that the velocity anisotropy of the BHB population is growing with
  distance from the Galactic centre.}
\label{fig:vellips}
\end{figure}

\begin{figure}
\includegraphics[width=\columnwidth]{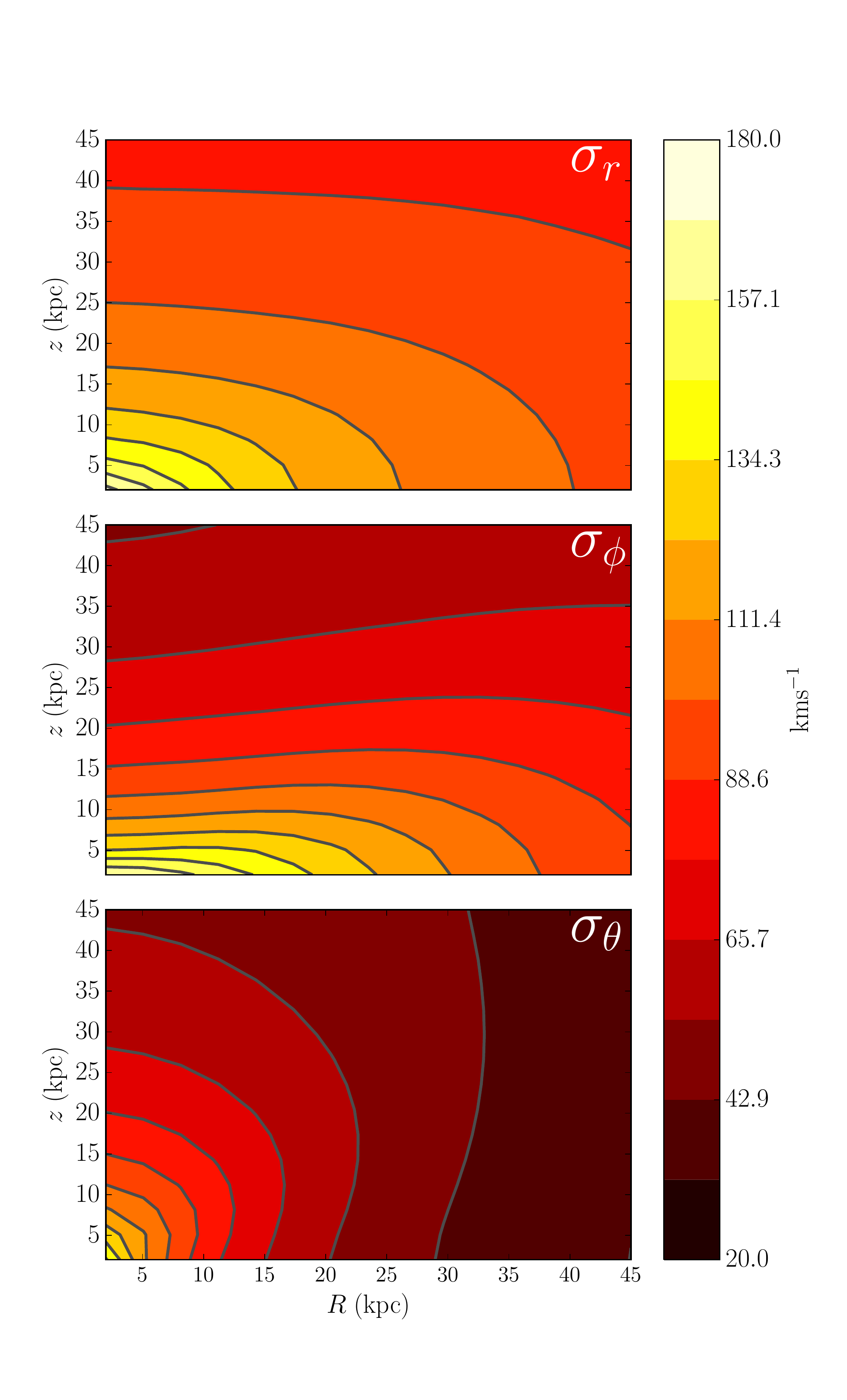}
\caption{Maps of the velocity dispersions $\sigma_r$ (top),
  $\sigma_\phi$ (middle) and $\sigma_\theta$ (bottom). The radial and
  azimuthal velocity dispersion contours are oblate, whereas the
  latitudinal velocity dispersions are prolate. The radial velocity
  dispersions decrease the least with Galactocentric distance, and the
  latitudinal dispersions the most -- as we expect for a flattened
  density.}
\label{fig:dispmaps}
\end{figure}

\subsection{Kinematics of the Stellar Halo}

We now study the kinematics of our best--fit model. Figure
\ref{fig:vellips} depicts the velocity ellipsoids of our model over a
range of positions in the Galaxy.  Our model naturally provides a
velocity ellipsoid that is everywhere aligned with spherical
coordinates, in agreement with the observations (e.g. \citealt{Sm09b},
\citealt{Bo10}).  The ellipsoid always has a long--axis pointing
towards the Galactic centre, meaning that the orbital distribution is
everywhere radially biased. The ellipsoid becomes more elongated with
distance from the Galactic centre, implying that the stellar halo
becomes more radially biased in the outskirts. Each ellipse is
coloured by the value of the anisotropy parameter
\begin{equation}
\beta(r) = 1 - \dfrac{\sigma _\phi^2 + \sigma_\theta^2}{2\sigma_r^2},
\end{equation}
which is a measure of the local velocity anisotropy. If $\beta>0$
($<0$), then the model is radially (tangentially) biased, and
$\beta=0$ implies an isotropic model. Figure \ref{fig:dispmaps}
depicts maps of the three spherical velocity dispersions with
position. We can see that the radial and azimuthal dispersions are
oblate, whereas the latitudinal dispersions are prolate in their
distribution. The latitudinal dispersions also exhibit a `pinching' in
their profile in the equatorial plane. This feature appears generic in
radially biased models that are oblate or triaxial \citep{Sa15c}.

\begin{figure}
\includegraphics[width=\columnwidth]{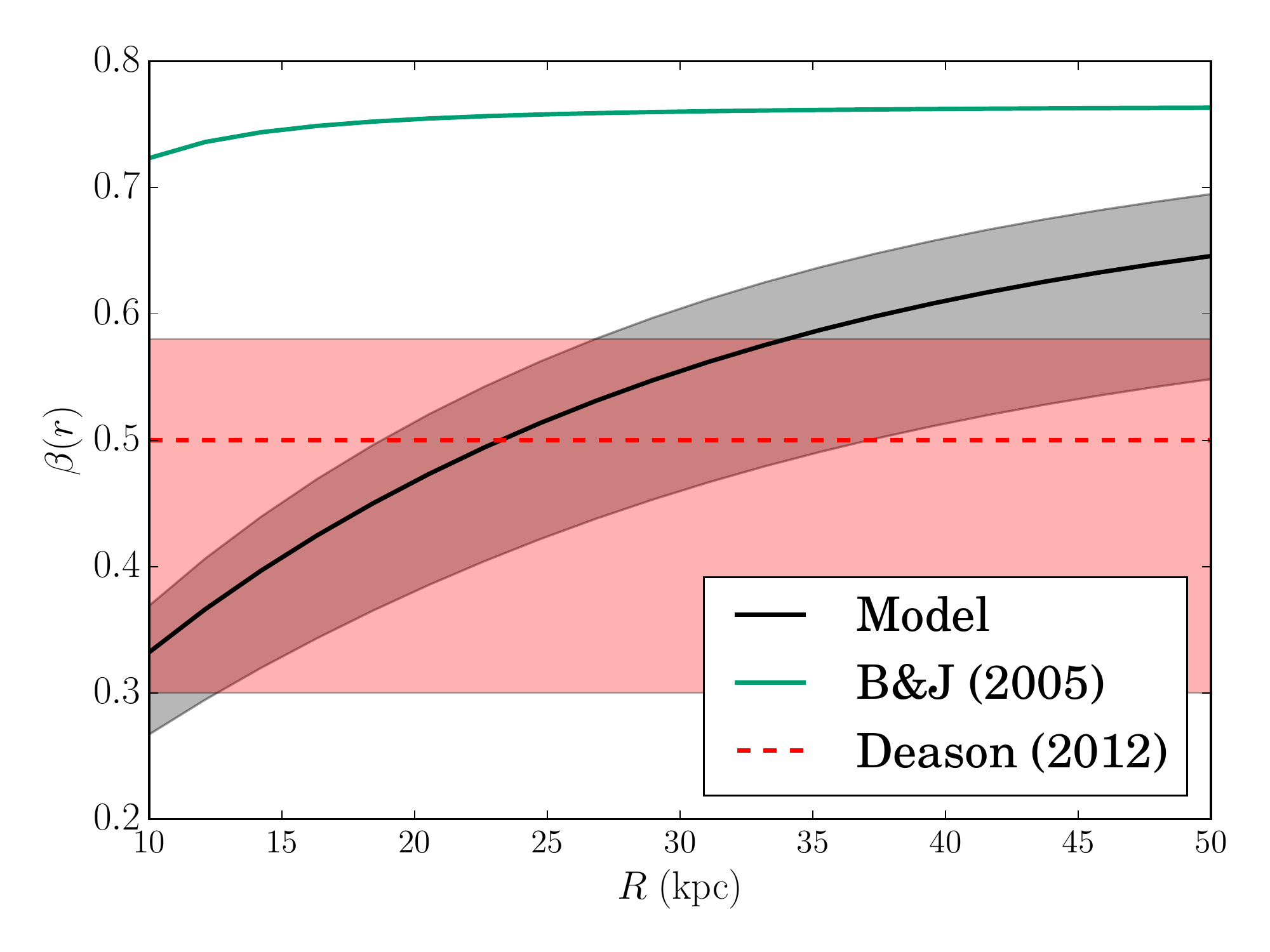}
\caption{Comparison between the anisotropy predicted by our model
  (black line, grey shade in 68\% confidence intervals), the best--fit 
  value from D12 (dashed red line, red shade in 68\% confidence intervals) 
  and the Bullock \& Johnston simulations 
  (green line). Our more sophisticated model is in agreement with the 
  analysis of D12 at the 68\% confidence level, but inconsistent with the simulated haloes.  The simulations 
  are far more radially biased than our model, implying that there may be significant
  differences between the Milky Way stellar halo formation history and
  the typical picture assumed in the $\Lambda$CDM paradigm.}
\label{fig:beta_profiles}
\end{figure}

An obvious comparison to make is between the prediction of D12, whose
DF enforced a globally constant anisotropy, and that of our
model. Figure \ref{fig:beta_profiles} depicts the anisotropy parameter and its 
uncertainty as a function of distance along the direction $R=z$ in our model, as well
as the value picked out by D12 and their confidence intervals. The two 
profiles are in agreement within the error bars at all positions for which 
we have data. Note too that D12's choice of
DFs was questioned by \citet{Fe13}, but our analysis seems to
vindicate the work.

Our model predicts an increasing anisotropy profile with radius, but
given the sparsity of the current data, it is difficult to assess how
realistic this is, so we turn to numerical simulations to make a
comparison. \citet{Bu05} simulated 11 different stellar haloes in the
$\Lambda$CDM context, each with the same dark halo mass and stellar disk
at redshift zero, but with distinct accretion histories. The stellar
haloes are built up over cosmic time by the accumulation of many
different subhaloes which have had stars `painted' onto
them. \citet{Ka12} then used the {\sc Galaxia} code \citep{Sh11} to
construct synthetic BHB populations for these simulations, and
analysed their velocity anisotropy profiles. They find that the mean
anisotropy profile of the 11 different BHB populations is well
represented by a function
\begin{equation}
\beta(r) = \dfrac{\beta_0r^2}{r^2+r_0^2},
\label{eq:betafunc}
\end{equation}
where $\beta_0=0.765$ and $r_0=2.4\kpc$. A rising anisotropy profile
is therefore consistent with their findings, although there are
significant differences between the simulation properties and those we
infer here. We instead find that the anisotropy profile of our model
is well represented by the function
\begin{equation}
\beta(r) = \beta_0 + \dfrac{(\beta_1 -\beta_0)r}{r+r_0},
\label{eq:betafunc1}
\end{equation}
with $\beta_0 = 0.05$, $\beta_1 = 0.82$ and $r_0 = 18.2\kpc$. The anisotropy profile
rises at a much lower rate with radius in our model than in the
simulations, only reaching $\beta \sim 0.7$ at $r \sim 100\kpc$. For reference, we also 
plot the anisotropy profile from the simulations in Figure \ref{fig:beta_profiles}. The
differences are curious, since our model is capable of creating
profiles similar to that of Equation (\ref{eq:betafunc}), but they do
not seem to be preferred. The \citet{Bu05} simulations involve
many small haloes accreting across cosmic time to build up the stellar
halo, but this is by no means the only possibility for the formation
history of the Galaxy's halo.

%It has been proposed that a single
%merger with a very massive dwarf galaxy could be responsible for the
%majority of the stars in the stellar halo, which could produce very
%different kinematics to those seen in other simulations. The stellar
%motions would possess memory of the progenitor's orbital eccentricity,
%for example. Different infall trajectories would produce very
%different anisotropy profiles, and only a small portion of the
%available initial conditions could create orbital distributions as
%radially biased as those seen in the Bullock \& Johnston
%simulations. Another possibility is that the observed BHB population
%was formed in--situ, which could also explain their apparently very
%smooth structure \citep{De11c}.

\begin{figure}
\includegraphics[width=\columnwidth]{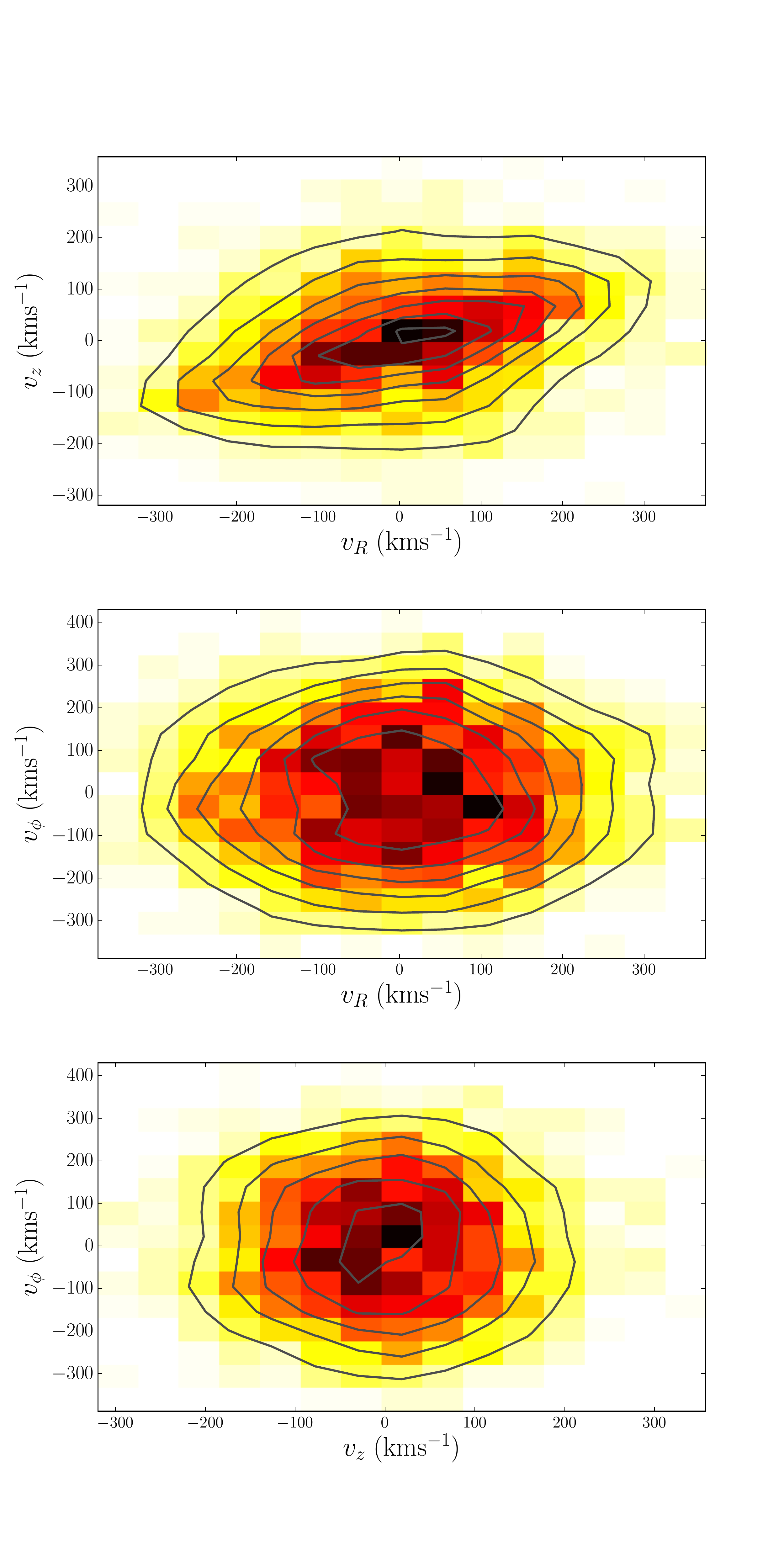}
\caption{The local cylindrical velocity distributions from a mock catalogue at the location $R=8\kpc$, $z=3.5\kpc$. 
We use a simple rejection sampling method to generate a catalogue from our model. Note the tilt of the velocity distribution 
in the $v_R-v_z$ projection, which results from the spherical alignment of the velocity ellipsoid.}
\label{fig:localdists}
\end{figure}

In the future, surveys such as Gaia will provide the three dimensional
velocity distributions of halo stars. In fact, such datacubes are
already beginning to be available. \citet[][hereafter B10]{Bo10} analysed the
motion of nearby stars from the SDSS using radial velocities coupled
with proper motions derived from SDSS and Palomar Observatory Sky
Survey (POSS) astrometry.  Figure \ref{fig:localdists} depicts the
two--dimensional velocity distributions of a mock catalogue of 2000
stars drawn from the best--fit model at the position $R=8\kpc$,
$z=3.5\kpc$. We draw the mock sample using a rejection sampling
method, with a sampling distribution that is the product of three
broad Gaussians in velocity space. The distributions are qualitatively
very similar to those found by B10, although there are some
differences (see their Figure 12).  The halo distributions presented
by B10 are broader in $v_z$, which is somewhat surprising given that
the halo is a flattened system. Furthermore, B10 infer a greater
velocity anisotropy in this region, $\beta \sim 0.65$, as compared
with our model, which has $\beta \sim 0.4$. This is perhaps suggestive
that the distribution of halo stars is significantly affected by the
presence of the disk, which is not accounted for in our
model. However, we have not convolved the velocities here with errors,
or considered any contamination from disk stars, both of which will
have an effect on the observed distributions. 

\section{Conclusions}

We have presented a new model for the Milky Way stellar halo, and used
it to fit a sample of Blue Horizontal Branch stars from the Sloan
Digital Sky Survey \citep{Xu11}. For the first time, we have used a
distribution function formulated in terms of the action integrals in
the fitting analysis.  The density profile generated by this DF is
flexible, which allows us to use the results on the SDSS photometry
from \citet{De11c} as a prior on the spatial distribution of our BHB
dataset. We argue that such a prior is necessary due to the unknown
selection function of the data. We then fit three model parameters
using the sample of 3534 stars, two corresponding to a simple
power--law gravitational potential and one that controls the
kinematics of the model.

Our results are consistent with a very flat rotation curve for the
Milky Way galaxy, with a mass enclosed at $50\kpc$ of $4.5 \times
10^{11} M_\odot$. This is in good agreement with other recent
estimates in the literature. The spherically aligned velocity
ellipsoid of our model is everywhere radially elongated, with a radial
bias that increases with Galactocentric distance as
\begin{equation}
\beta(r) = \beta_0 + \dfrac{(\beta_1 -\beta_0)r}{r+r_0},
\label{eq:betafunc1}
\end{equation}
with $\beta_0 = 0.05$, $\beta_1 = 0.82$ and $r_0 = 18.2\kpc$. This behaviour
qualitatively agrees with results seen in numerical simulations, in
that the anisotropy parameter is an increasing function of radius, but
our model suggests a much gentler growth of $\beta$ than in the
simulations. Given that the model can produce far more radially biased
distributions than our maximum--likelihood model, it is interesting
that the data do not seem to be consistent with the canonical
simulations of stellar halo formation. This is suggestive that the
formation history of the stellar halo may be somewhat different from
that implied by the models of \citet{Bu05}.

Though our model is the most complex DF yet fitted to data on the
stellar halo, it still has some obvious limitations. We have tightly
constrained the density profile of the stars in our model so that it
is in agreement with past analysis of the stellar halo, a step we
unfortunately believe to be necessary owing to the unknown selection
function of the data (c.f. Das \& Binney, in prep.). Evidently, in applications
to datasets provided by Milky Way surveys such as Gaia, we will have a
better understanding of the selection function and be able to
constrain the spatial and kinematic structure of the stars, as well as
the Galactic potential.

Another layer of sophistication can be added to our models in order to
account for rotation in the stellar halo. Here, we have presumed that
there is no net rotation of the halo, when in reality this score is by
no means settled ~\citep[e.g.,][]{De11a,Fe13}. One way of
introducing a mean azimuthal streaming velocity is by modifying the DF
to become ~\citep[e.g.,][]{De11a,Bi14}
\begin{equation}
f_\mathrm{rot}(\acts) = \left[\gamma\mathrm{tanh}\left(J_\phi/\delta J\right) + (1-\gamma)\right]f(\acts),
\end{equation}
where $\delta J$ is a small action to avoid discontinuities in the
gradient of the DF and $f(\acts)$ is the original ansatz from Equation
(\ref{eq:ourDF}). The parameter $\gamma$ then takes values between $-1/2$
(maximal retrograde motion) and $1/2$ (maximal prograde rotation).  We can then add the
extra parameter $\gamma$ to the analysis and search for rotation in
the stellar halo.

The model presented here is constructed in a very simple potential,
meaning that it is possible to write down an ansatz for a DF with
predictable features. However, given the rapid recent developments on
action estimation in generic potentials (\citealt{Bi12a},
\citealt{Bo14}, \citealt{Sa14a}, \citealt{Sa15a}), it is now of
importance that we develop flexible DFs with well--understood
properties in more realistic Galactic potentials. Current models of
spheroidal components of galaxies (\citealt{Wi15}, \citealt{Po15})
have been developed to produce certain features in spherical
potentials, and even though they can be modified for use in
axisymmetric potentials (e.g. Das \& Binney, in prep.), their properties are less well
understood in this case. Eventually, it will be the case that we
simultaneously fit multiple--component Galaxy models
(e.g. \citealt{Pi15}) to data from various surveys, but for this to be
an effective procedure we must ensure that the DFs used for the
various components are sufficiently sophisticated to model the
high-quality data that is to appear over the next few years
\citep{Pe01}. For example, \citet{Sa15c} investigate triaxial
generalisations of the models in \citet{Wi15}, which should provide
valuable information as to how to progress can be made in this area.

\section*{Acknowledgments}

AAW thanks the Science and Technology Facilities Council (STFC) for
the award of a studentship. We thank Vasily Belokurov, Payel Das, Alis
Deason, and Jason Sanders for some useful conversations, and Simon
Gibbons for generously sharing his code and numerical expertise. We also 
thank the referee for a careful report that made several aspects of the paper 
much clearer.

\bibliography{stellar_halo}
\bibliographystyle{mn2e}

\end{document}

%% file: defn.tex
\newcommand{\kpc}{\rm\thinspace kpc}